\documentclass{kapproc} 

\usepackage{procps}

\usepackage[dvips]{graphicx}

\chaptersection
\upperandlowercase

\begin{document}

\kluwerbib

\articletitle[The XMM-2df Cluster Survey]
{The XMM-2df Cluster Survey\\
}

\author{Gaga, T.$^{1,2}$, Plionis, M.$^{1,3}$, Georgantopoulos, I.$^{1}$, Georgakakis, A.$^{1}$,
               Basilakos, S.$^{1}$, Stewart, G.C.$^{4}$, Kolokotronis, V.$^{1}$, Stobbart,
               A.M.$^{4}$\\
               }

\affil{$^{1}$ Institute of Astronomy \& Astrophysics, National
Observatory of Athens \\ I. Metaxa \& V. Pavlou, Athens, 15236,
Greece}

\affil{$^{2}$ Section of Astrophysics, Astronomy and Mechanics,
Department of Physics, University of \\ Athens, Panepistimiopolis,
GR-157 83, Zografos, Athens, Greece}

\affil{$^{3}$ Instituto Nacional de Astrofisica, Optica y Electronica (INAOE) Apartado Postal 51 y 216,\\ 72000, Puebla, Pue., Mexico}

\affil{$^{4}$ Department of Physics and Astronomy, University of
Leicester,Leicester LE1 7RH, UK} \email{dg@astro.noa.gr}

\begin{abstract}
We present the results from a shallow (2-10 ksec) XMM/2dF survey.
Our survey covers 18 XMM fields ($\sim 5 {\rm deg}^2$) previously
spectroscopically followed up with the Anglo-Australian telescope
2-degree field facility. About half of the fields are also covered
by the Sloan Digital Sky Survey (SDSS). We are searching for
extended sources using the XMM SAS maximum likelihood algorithm in
the 0.3-2 keV band and we have detected 14 candidate clusters down
to a flux of $\sim2\times10^{-14} cgs$. Our preliminary results
show that: i) the redshift distribution peaks at relatively high
redshifts ($\sim0.4$) as expected from the Rosati et al.
$\Phi(L)$, ii) some of our X-ray clusters appear to have optical
counterparts.
\end{abstract}


\section{Cluster Detection in the XMM fields and Results}
We have analyzed 18 XMM-fields located near the South Galactic Pole
(SGP; Ra = $00^{h}$ $57^{min}$, Dec = $-28^{\rm\circ}$
$00^{\rm\prime}$; total of 9 fields) and in the North Galactic
Pole (NGP; Ra = $13^{h}$ $41^{min}$, Dec = $00^{\rm\circ}$
$00^{\rm\prime}$; total of 9 fields), covering an area of $\sim5
{\rm deg}^2$ and with exposure time between 2 and 10 ksec.
 We have excluded 4 southern fields and 1 northern
field because of the high particle background.

In order to detect extended emission we use the wavelet-based
XMM-SAS source detection task EWAVELET, with a detection threshold
of $5\sigma$, in combination with the EMLDETECT task with an extension
probability >0.995. We detect: i) 5 candidate southern clusters in
the MOS1 and MOS2 mosaic and 3 extended sources in the PN (out of
which 3 overlap with the MOS detections) and ii) 6 candidate
northern clusters in the MOS1 and MOS2 mosaic, 1 in MOS1 only and
5 extended sources in the PN (out of which 3 overlap with the MOS
detections). The faintest extended source has a flux of $\sim$ 2
$\times$ 10$^{-14}$ cgs, as estimated by the EMLDETECT task.
Visual inspection suggests that the small overlap between the PN
and the MOS detected cluster candidates can be attributed to (i)
the presence of gaps in the PN and (ii) the elevated particle
background of this detector.

For the NGP cluster candidates we compare the photometric redshift
distribution (from SDSS) within a circular region of
$1^{\rm\prime}$ - $3^{\rm\prime}$ centered on the X-ray position
with that of the field. This preliminary analysis suggests an
overdensity of optical galaxies at photometric redshifts z $\sim$
0.4 in excess to the expectations from the Rosati et al 2002
luminosity function.

In the figures we present the XMM EPIC (mosaic of MOS1 and MOS2)
contours of 2 of the candidate clusters, overlaid on the
optical images, using the DSS and the SDSS for the southern and
northern fields respectively.

We fitted to each cluster a King profile using fixed values for
the $\beta$ parameter ($\beta=0.7$ and $\beta=1.0$). We then
estimated the flux of each cluster by integrating their King
profile to infinity and we found that the lower cluster luminosity is
$\sim10^{42} \; cgs$. Varying the $\beta$ parameter between 0.7
- 1 translates to a $\sim30\%$ change in $R_{\rm core}$. We also found 
that the range of the cluster temperatures is between 1-3 keV.

To test the reliability of our procedure we have performed
simulations of the expected cluster detection on XMM EPIC, using
the Rosati et al (2002) $\Phi(L)$ and the SAS simulator (SciSim). We
find that we should expect roughly $\sim1.5$ clusters per field
with f $>$ 2 $\times$ 10$^{-14}$cgs, which is in  rough agreement
with our number of detected clusters.

\begin{figure}[t]

%\centerline{\hbox{\includegraphics[width=5.cm]{gaga_f1.ps}
%\includegraphics[width=5.4cm]{gaga_f2.ps}}}

{\caption{In the above figures the optical DSS image (for the Northern
cluster at the left) and the SDSS image (for the southern cluster
at the right) is overlaid with the X-ray contours}
}
\end{figure}


\begin{chapthebibliography}{1}
%\bibitem{goto}
%Goto, J. et al (2002), AJ, 123, p.1807.

\bibitem{rosati}
Rosati, P. et al (2002), Annual Review of Astronomy and
Astrophysics, vol.40, p.539 - 577.

\end{chapthebibliography}
\end{document}